\def\lesssim{\,\lower2truept\hbox{${<\atop\hbox{\raise4truept\hbox{$\sim$}}}$}\,}
\def\gtrsim{\,\lower2truept\hbox{${>\atop\hbox{\raise4truept\hbox{$\sim$}}}$}\,}

\def \SAIT #1 #2 {{\em Mem.\ Soc.\ Astron.\ It.\/} {\bf #1}, #2}
\def \MESS #1 #2 {{\em The Messenger\/} {#1}, #2}
\def \ASTRNACH #1 #2 {{ Astron. Nach.\/} { #1}, #2}
\def \AAP #1 #2 {{ A{\rm \&}A\/} {#1}, #2}
\def \AAL #1 #2 {{ A{\rm \&}A\/} {#1}, L#2}
\def \AAR #1 #2 {{ A{\rm \&}AR\/} {#1}, #2}
\def \AAS #1 #2 {{ A{\rm \&}AS\/} {#1}, #2}
\def \AJ #1 #2 {{ AJ\/} {#1}, #2}
\def \ANNREV #1 #2 {{ ARA{\rm \&}A\/} {#1},#2}
\def \APJ #1 #2 {{ ApJ\/} {#1}, #2}
\def \APJL #1 #2 {{ ApJ\/} {#1}, L#2}
\def \APJS #1 #2 {{ ApJS\/} {#1}, #2}
\def \APSS #1 #2 {{ Ap{\rm \&}SS\/} {#1}, #2}
\def \ASR #1 #2 {{ Adv. Space Res.\/} {#1}, #2}
\def \BAIC #1 #2 {{ Bull. Astron. Inst. Czechosl.\/} { #1}, #2}
\def \JSQRT #1 #2 {{ J. Quant. Spectrosc. Radiat. Transfer\/} {#1}, #2}
\def \MN #1 #2 {{ MNRAS\/} { #1}, #2}
\def \MEM #1 #2 {{ Mem. R. Astr. Soc.\/} { #1}, #2}
\def \PLR #1 #2 {{ Phys. Lett. Rev.\/} { #1}, #2}
\def \PASJ #1 #2 {{ Publ. Astron. Soc. Japan\/} { #1}, #2}
\def \PASP #1 #2 {{ Publ. Astr. Soc. Pacific\/} { #1}, #2}
\def \NAT #1 #2 {{ Nat\/} { #1}, #2}
\def \ACTqA #1 #2 {{ Acta Astron.\/} { #1}, #2}
\documentclass[]{aa}
\usepackage{graphicx}

    \def\smallskip{\vskip 6pt}
    
    \def\M12{${\rm M_{12}}$}

\begin{document}

\title{Modelling the Spectral Energy Distribution of Compact Luminous Infrared
Galaxies: Constraints from High Frequency Radio Data}

\author{O.R. Prouton$^1$, A. Bressan$^{2,3}$, M. Clemens$^{1}$,
    A. Franceschini$^{1}$, G.L. Granato$^{2,3}$, L. Silva$^{4}$}

\institute{Dipartimento di Astronomia, Universit\`a degli Studi di
    Padova, Vicolo dell'Osservatorio 2, I-35122 Padova, Italy
    \and
    INAF, Osservatorio Astronomico di Padova, Vicolo
    dell'Osservatorio 5, I-35122 Padova, Italy
            \and
   SISSA, Strada Costiera, I-34131 Trieste, Italy
               \and
   INAF, Osservatorio Astronomico di Trieste, Via Tiepolo 11, I-34131 Trieste, Italy}

\offprints{A. Bressan, \email{bressan@pd.astro.it}}

\date{Received  / Accepted 16/03/2004}


\abstract{
We have performed 23 GHz VLA observations of 7 compact, luminous
infrared galaxies, selected to have evidence of starburst
activity. New and published multi--frequency data are combined to
obtain the spectral energy distributions of all 7 galaxies from the
near--infrared to the radio (at 1.4 GHz). These SEDs are compared with
new models, for dust enshrouded galaxies, which account for both
starburst and AGN components.

In all 7 galaxies the starburst provides the dominant contribution to
the infrared luminosity; in 4 sources no contribution from an AGN is
required. Although AGN may contribute up to 50 percent of the total
far--infrared emission, the starbursts always dominate in the
radio. The SEDs of most of our sources are best fit with a very high
optical depth of $\sim 50$ at $1\;\rm \mu m$.

The scatter in the far--infrared/radio correlation, found among
luminous IRAS sources, is due mainly to the different evolutionary
status of their starburst components. The short time--scale of the
star formation process amplifies the delay between the far--infrared
and radio emission. This becomes more evident at low radio frequencies
(below about 1 GHz) where synchrotron radiation is the dominant
process. In the far--infrared (at wavelengths shorter than 100 $\mu$m)
an additional source of scatter is provided by AGN, where present. AGN
may be detected in the near--infrared by the absence of the knee,
typical of stellar photospheres. However, near--infrared data alone
cannot constrain the level at which AGN contribute because the
interpretation of their observed properties, in this wave--band,
depends strongly on model parameters.

}

\titlerunning{Modelling the SED of Compact Luminous Infrared Galaxies}

\authorrunning{O.R. Prouton et al.}

\maketitle

\section{Introduction}\label{introduction}

One of the most intriguing aspects of IRAS galaxies 
 (Soifer et al., 1984, Sanders \& Mirabel 1996), 
characterized by their extremely cool emission (peaking
between 60 and $100\; \rm \mu m$), is the origin of their high
luminosities, often exceeding $10^{12}\; \rm L_\odot$.  
In such galaxies it is clear that we observe radiation from dust at temperatures of
a few 10s of Kelvin. However, the question of whether the heat source is
baryonic nuclear processing in young stars, or accretion of baryons onto a
central black hole is still debated.
  Recent evidence (e.g. Ferrarese 2002) that all spheroids harbour a central black hole, suggested by
the existence of a tight correlation between the masses of black holes and the
spheroidal components, implies that there may be a relationship between these two processes.
 The
evidence that almost all ultra luminous IRAS sources (ULIRGs) exhibit signatures
of on-going or recent interaction, suggests that the triggering mechanism is
dynamical interaction. 

Several questions remain:
\begin{enumerate}
\item What are the relative fractions of mass going into star formation and
  accretion respectively? 
\item How tight is the link between star formation and nuclear activity? And
  is there any causal link between the two processes? 
\item How important is the role of mutual feedback in the termination of these processes?
\end{enumerate}
 The answers to these questions 
 take on greater importance if, as deduced by deep
ISO and sub--millimetric surveys (e.g. Franceschini et al., 2001, 2003; Elbaz
et al., 2002; Smail et al., 2000; Barger et al., 2000), ULIRGs are the local counterparts of a
larger population of luminous galaxies at redshifts z$\geq 0.5$.

Recent spectroscopic
determinations of the redshift of SCUBA sources (Smail et al., 2003) 
confirm that these objects have extreme luminosities. If these luminosities
were due to starburst episodes (accounting for possible
magnification by gravitational lensing), they would correspond to star formation rates (SFRs) in
excess of several thousand M$_\odot$/yr. 
If these extreme SFRs were maintained over a dynamical
timescale, a spheroidal component of a typical L$_*$ galaxy could be assembled.

Both star formation and nuclear activity can be significant sources of energy
feedback into the interstellar and intergalactic media -- galactic winds and
re-ionization respectively (e.g. Granato et al., 2001; Granato et al., 2003). The latter mechanism is one of the most
important, if least understood, ingredients in current models of galaxy
formation (Cavaliere, Lapi \& Menci, 2002).

Bressan, Silva \& Granato (2002) showed that comparison of the radio
and far--infrared (FIR) emission may provide an accurate
diagnostic for the study of the star formation history during the
strong starburst phase characteristic of compact ULIRGs.  By means of
new starburst models they were able to show that deviations from the
FIR/radio correlation could be used to derive the evolutionary status
of the starburst.  These deviations are expected in bursts of short
duration because at early times (a few $10^6\;\rm y$) even the most
massive stars formed in the burst will not have ended their lives as
supernovae and no excess synchrotron emission should result from the
starburst activity. Such starbursts would be relatively strong FIR emitters. Another important signature
of a young starburst is its radio slope; as young starbursts are not yet
dominated by synchrotron radiation from supernovae their radio
slope  should be flatter than $S_{\nu}\propto \nu^{-0.7}$, due to the large free-free component from ionized gas.

Bressan, Silva \& Granato's (2002) ability to derive the radio slope of
objects in Condon's (1991) sample of compact ULIRGs 
 was limited by their use of existing radiometry at $1.4\;\rm GHz$ 
 (which may be affected by free-free absorption). Moreover, they
did not attempt to perform a complete fit to the complete SED of the
galaxies, nor did they  account for the effects of the AGN in the
infrared.

 The new $23\; \rm GHz$ fluxes presented here 
allow for both the determination of the radio spectrum, unaffected by free-free 
absorption, and 
place a useful constraint on the level of free-free emission.
The combination of radio data with IRAS fluxes at 60 and
$100\;\rm \mu m$ provides a strong constraint on the age of starburst
activity. FIR data below $60\;\rm \mu m$ are sensitive to the
contribution of AGN, which may then be estimated as the
excess above the starburst component.

By studying local ULIRGs we can probe all of the important physical mechanisms
involved in the formation of spheroidal systems: dynamical interaction; the
triggering of star formation; QSO activation and, perhaps most important,
feedback. In this paper we present new $23\;\rm GHz$ observations of 7 
compact ULIRGs. With these new data and existing archival photometry/radiometry, we
study their star formation history and possible AGN contributions.

Source selection is briefly discussed in Section
\ref{sample}, corresponding observations are presented in Section
\ref{observations}. In Section \ref{datared} we describe the procedure used 
 to obtain the flux measurements of the observed sources while  in
Section \ref{results} we present the data.  In Section \ref{sed} we
present starburst and AGN models adopted to fit the SEDs of the
observed galaxies. Discussion of the results and  conclusions are
given in Sections \ref{discussion} and \ref{conclusions},
respectively.

\begin{table*} \caption{Previously published radio, FIR, mid-IR and sub--mm fluxes of the
compact luminous infrared galaxy 
sample.}
\begin{tabular}{lccccrrrrcc}
\hline
\hline
Source   & $\rm{S}_{1.4}$ &  $\rm{S}_{4.85}$ &$\rm{S}_{8.44}$ &$q$&$\rm{S}_{12\mu}$&$\rm{S}_{25\mu}$&$\rm{S}_{60\mu}$&$\rm{S}_{100\mu}$
     & $\rm{S}_{850\mu}$ & log L$_{\rm FIR}$ \\
Name     & / mJy      & / mJy        &  / mJy         &
     -  & /mJy & /mJy & /mJy & /mJy & /mJy & L$_{\odot}$ \\
     & (1)        & (2)              &  (3)       &
     (4)  & (5) & (6) & (7) & (8) & (9) & (10)  \\
\hline
NGC0034        &$67.5\pm2.5$   & \ldots  & $15.2\pm0.8$                &2.52 &360    &2380 &16080 &16970 &\ldots    & 11.54\\
CGCG436-30     &$50.2\pm1.6$   & 21.5    & $12.7\pm0.6$                &2.46 &270    &1410 &10720 &9600  &$39 \pm 8$& 11.77\\
IRAS01364-1042 &$15.8\pm0.7$   & \ldots  & $8.2\pm0.4$                 &2.67 &$<150$ &430  &6530  &7000  &\ldots    & 11.93\\
IRAS08572+3915 &$4.89\pm0.19^*$& \ldots  & $4.1\pm0.2$                 &3.11 &340    &1840 &7660  &5060  &$17\pm7$  & 12.24\\
NGC7469        & $180.5\pm5.4$ & $70\pm8$&$>15.0\pm0.8^\dagger$        &2.29 &1488   &5430 &25740 &32466 &$264\pm30$& 11.68\\
IC5298         & $34.7\pm1.4$  & 13.9    & $8.2\pm0.4$                 &2.65 &288    &1690 &7880  &10480 &$77\pm15$ & 11.61\\
UGC12812       & $70.7\pm2.2$  & 31.1    &$21.5\pm1.1^{\dagger\dagger}$&2.51 &492    &2470 &16710 &20130 &$132\pm25$& 11.54\\
\hline
\end{tabular}
{\footnotesize
(1):  NVSS 1.4 GHz flux densities Condon et al. (1998); $^*$ FIRST 1.4 GHz flux densities Becker et al. (1995);  (2): Sopp \& Alexander (1992);  (3): Condon et al.'s (1991) total 8.44 GHz source flux densities, $^\dag$ main component flux density, $^{\dag\dag}$ indicates a relatively uncertain flux density due to surrounding diffuse emission. Errors in the 8.44 GHz fluxes were estimated from the observational parameters given in Condon et al. using the method described in Section \ref{radiometrysection}  (4): logarithmic far infrared to 1.4 GHz luminosity ratio as defined by Condon et al. (1991);  (5), (6), (7) \& (8): IRAS-BGS flux densities (Soifer et al., 1989); (9): SCUBA $850\;\rm \mu m$ flux densities Dunne et al. (2000); (10) far infrared luminosity in the range $8-1000\;\rm \mu m$.
}
\label{sourceproperties}
\end{table*}
\section{Sample Selection}
\label{sample}

Sources were selected from Condon et al.'s (1991) catalogue of compact
ULIRGs. These, in turn, are a sub--set of sources in the Revised IRAS Bright Galaxy
Survey (Soifer et al., 1989).
Objects in our sample were selected to have a
ratio of FIR to $1.4\;\rm GHz$ luminosity $q \geq 2.5 $.
Such a ratio should, according to the starburst models of Bressan, Silva \& Granato
(2002), help ensure emission is dominated by a young starburst. All sources were
required to be bright enough, at 8.44 GHz, to ensure a strong detection
(estimated to be in excess of $7 \sigma$) within a few minutes integration and
to be grouped close enough in R.A. to enable the entire sample to be observed
within a few hours at the VLA. NGC 7469
($q=2.29$) was added as a well studied comparison source, which is known to
contain a central AGN and a circumnuclear starbursting ring.

Integrated radio, FIR and sub-mm fluxes of the sources
are presented in Table \ref{sourceproperties}. Flux densities at $1.4\;\rm
GHz$ were taken from either the FIRST, NVSS or Condon et al. (1990)
catalogues, $4.85\;\rm GHz$ fluxes from Sopp \& Alexander (1992) and $8.44\;\rm GHz$
fluxes from Condon et al. (1991). Integrated near--infrared (NIR) fluxes from
the 2MASS archive (Jarrett et al., 2003) are
shown in Table \ref{tab_nir}.

\section{Observations}\label{observations}

K--band ($22.5\; \rm GHz$) observations were made with the VLA in
D--array on 2002 January $3^{\rm rd}$, 0800--0900 LST and 2002 January
$5^{\rm th}$, 2100--2400 LST. We used a bandwidth of $50\;\rm MHz$. During
our observations 18 antennae were fitted with new low--noise K--band
receivers, the remaining 9  with old receivers. The estimated point
source r.m.s. noise in a naturally weighted image was then $ 5.958
(t)^{-1/2}\ \rm{mJy \ beam}^{-1}$  where $t$ is the integration time
in seconds.  At $23 \; \rm GHz$ the angular resolution of D--array is
approximately 2.5 arcsec.

Pointing errors for the VLA antennae are known to significantly
degrade the sensitivity of the instrument at high frequencies. In
order to obtain the most accurate possible flux density measurements
we fine--tuned the pointing accuracy using the procedure termed second
order referenced
pointing\footnote{http://www.aoc.nrao.edu/vla/html/refpt.shtml}. At
the start of both observing runs the pointing offset was determined
for the flux calibrator in the sensitive X ($8.4\; \rm GHz$) band. The
second--order pointing correction was determined at the normal
observing frequency, $23\; \rm GHz$. This pointing offset was
applied to measurements of the phase calibrators and the same
procedure followed to find accurate pointing corrections near to each
source position. Finally these phase calibrator offsets were applied
when pointing at the target source.
Each source and
its associated phase calibrator was observed for 4 (IRAS 0136-10,
NGC 34, NGC 7469, IRAS 08572+39) or 5 (CGCG 436-030, UGC 12812, IC 5298)
snapshots.

The choice of flux density calibrators was dictated by the available
hour angle coverage: on January $3^{\rm rd}$ 3C147 (0542+498) and on January
$5^{\rm th}$ 3C48 (0137+331) were observed. Phase calibration sources were
chosen from the subset of the NRAO VLA calibrator
list\footnote{http://www.aoc.nrao.edu/$\sim$gtaylor/csource.html} with
positions known to better than 2 mas.  Each source was associated with
the nearest calibrator with tabulated 15 and $43\;\rm GHz$ (2 and 0.7 cm)
fluxes and designated {\emph P} or {\emph S} at both these
frequencies, none had tabulated $22\; \rm GHz$ (1.3 cm) flux
densities.  (For a {\emph P} class calibrator amplitude closure errors
are expected to be $\leq$ 3 percent; for an {\emph S} class calibrator
in the range 3 to 10 percent).
\begin{table} \caption{2MASS J, H and  K--band total fluxes of the
compact luminous infrared galaxy sample. }
\begin{tabular}{lccc}
\hline
\hline
Source   & J   & H   &  Ks  \\
Name     & /mJy& /mJy& /mJy \\
\hline
  NGC0034            &  50.8$\pm$1.2  & 66.8$\pm$1.8  &62.3$\pm$1.8\\
  CGCG436-30         &  14.4$\pm$0.51 & 18.3$\pm$0.65 &20.1$\pm$1.1\\
  IRAS01364-1042     &  4.11$\pm$0.35 & 5.13$\pm$0.48 & 5.13$\pm$0.58\\
  IRAS08572+3915$^*$ &  1.7$\pm$0.11 & 3.0$\pm$0.20 &3.9$\pm$0.26\\
  NGC7469            &  144  $\pm$3.3 & 205$\pm$ 4.8  & 193$\pm$5.2\\
  IC5298             &  33.3$\pm$0.9  & 45.2$\pm$1.4  &46.5$\pm$1.7\\
  UGC12812           &  56.9$\pm$0.9  & 72.3$\pm$1.4  &74.5$\pm$1.7\\
\hline
\end{tabular}
{\footnotesize\\
$^*$ From Spinoglio et al. (1995)}
\label{tab_nir}
\end{table}

\begin{table*} \caption{Assumed flux densities of flux calibrators (top) and
calculated flux densities of phase calibrators (bottom).}
\begin{tabular}{llcc}
\hline
\hline
Source & Calibrator  & S1 /mJy  & S2 /mJy   \\
\hline
$\rm{All}^*$ & 3C48 (0137+331)   & 1117.4   & 1120.2    \\
IRAS08572+39 & 3C147 (0542+498)  & 1794.9   & 1799.0    \\
\hline
NGC0034      & 2358-103      & $573.35 \pm 4.48 $ & $570.47 \pm 4.54$ \\
CGCG436      & 0121+118          & $1175.52 \pm 7.38$ & $1177.64 \pm 6.81$ \\ 
IRAS0136-10  &0141-094           & $739.52 \pm 3.42 $ & $739.57 \pm 3.46$ \\ 
IRAS08572+39 & 0927+390      & $7710.36\pm 149.26$ & $7665.30 \pm 147.39$ \\ 
NGC7469      & 2320+052      & $567.67 \pm 4.31 $ & $564.61 \pm 4.41$ \\ 
IC5298       & 2321+275      & $716.32 \pm 8.79 $ & $719.28 \pm 9.00$ \\ 
UGC12812     & 2358+199      & $268.75 \pm 2.19 $ & $267.78 \pm 2.21$ \\ 
\hline
\end{tabular}
{\footnotesize

$^*$: Except IRAS08572+39;\\
S1: Flux density in the $50\;\rm MHz$ band centered at $22.485\;\rm GHz$; \\
S2: Flux density in the $50\;\rm MHz$ band centered at $22.435\;\rm GHz$.}
\label{calibratorfluxes}
\end{table*}

\section{Data Reduction}\label{datared}

\subsection{Calibration \& Editing}

Rapid ionospheric/tropospheric phase instabilities, coupled with
angular resolution capable of resolving structure in the calibration
sources, demands particular care when reducing data in the $22\; \rm
GHz$ band. Calibration and editing were performed within the NRAO data
reduction package {\sc aips},  following the method suggested for high
frequency data obtained with the VLA (e.g. AIPS Cookbook Appendix D).

Data were loaded such that nominal sensitivities were used to give
relative weight to data points and opacity and gain curve corrections
were applied. Throughout calibration both data and their associated
weights were calibrated. To avoid the effects associated with
resolving the structure of the primary flux calibrators, clean
component models (available via the VLA
website\footnote{http://www.aoc.nrao.edu/vla/html/highfreq/}), were
used for both flux calibrators, 3C 147 and 3C 48, when solving for
antenna--based phase and amplitude calibration solutions. Calibration
solution intervals were made short (20 seconds) to account for rapidly
varying atmospheric phases.

\begin{table*} \caption{Measured $23\; \rm GHz$ source positions and flux
 densities at 2002.01.}\label{measuredfluxestable}
\begin{tabular}{lccccc}
\hline
\hline
Source          & RA   & Dec                & $r$  & $\phi$ & $S_{\rm uv}$ \\
                &\multicolumn{2}{c}{(J2000)}& /mas & /deg   & /mJy    \\
\hline
NGC0034     & 00 11 06.550 & -12 06 26.330 & 1424.07 &  179.587 & $7.41\pm0.14$ \\
CGCG436-30  & 01 20 02.722 & +14 21 42.940 & 1221.50 & -108.351 & $8.73\pm0.31$ \\
IRAS0136-10 & 01 38 52.870 & -10 27 11.700 & 108.37  &  -90.993 & $3.97\pm0.19$ \\
IRAS0857+39 & 09 00 25.390 & +39 03 54.400 & 419.781 & -63.642  & $3.18\pm0.30$ \\
NGC7469     & 23 03 15.623 & +08 52 26.390 & 346.625 & -177.478 & $17.50\pm0.50$ \\
IC5298      & 23 16 00.690 & +25 33 24.200 & 337.614 & -136.246 & $3.86\pm0.18$ \\
UGC12812    & 23 51 26.802 & +20 35 09.870 & 831.105 & -59.125  & $9.95\pm0.29$ \\
\hline
\end{tabular}
\label{sourcefluxes}
\end{table*}

\begin{figure*}[t]
{\centering \begin{tabular}{ccc}
\resizebox*{0.32\textwidth}{!}{\includegraphics{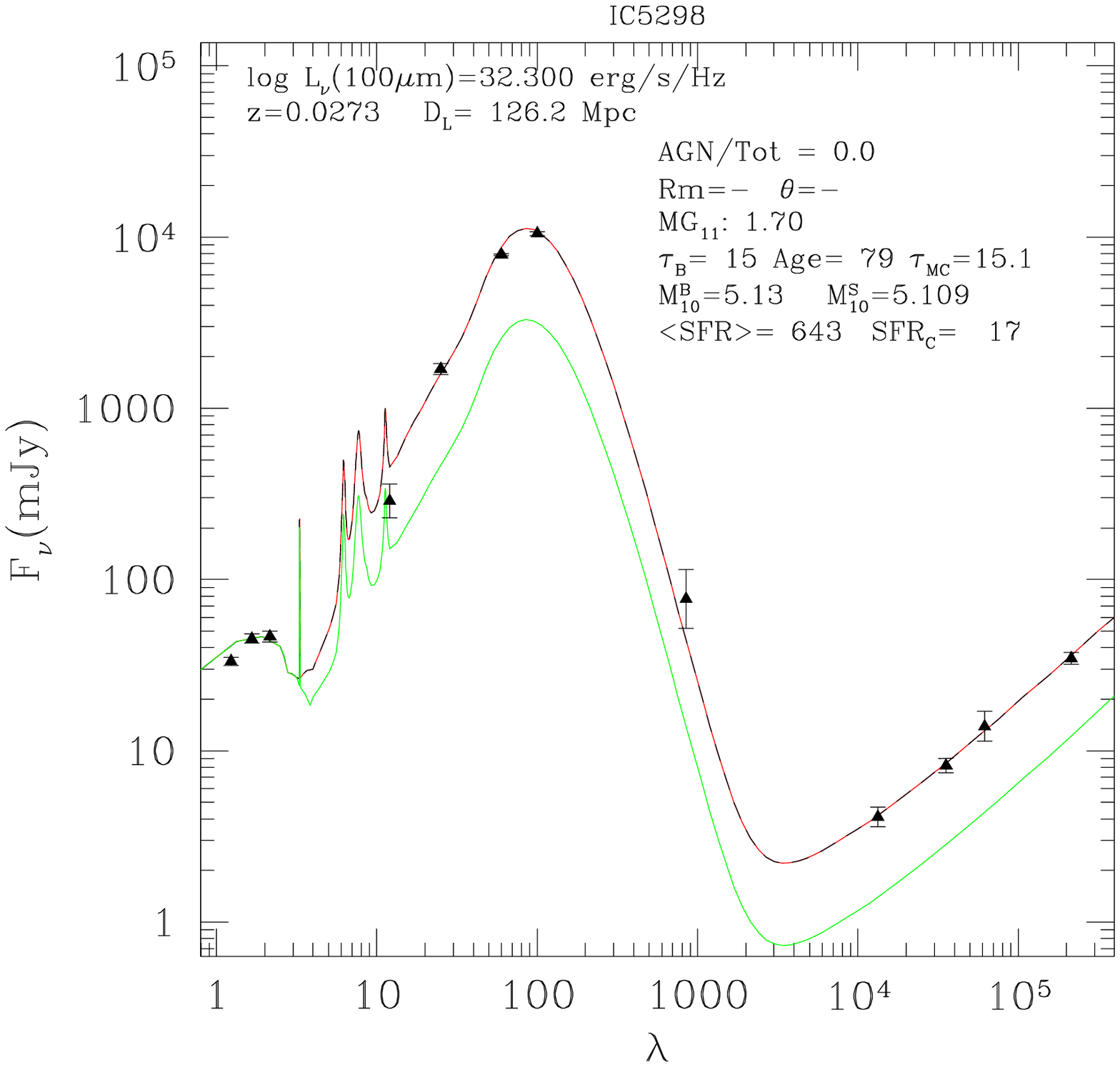}} &
\resizebox*{0.32\textwidth}{!}{\includegraphics{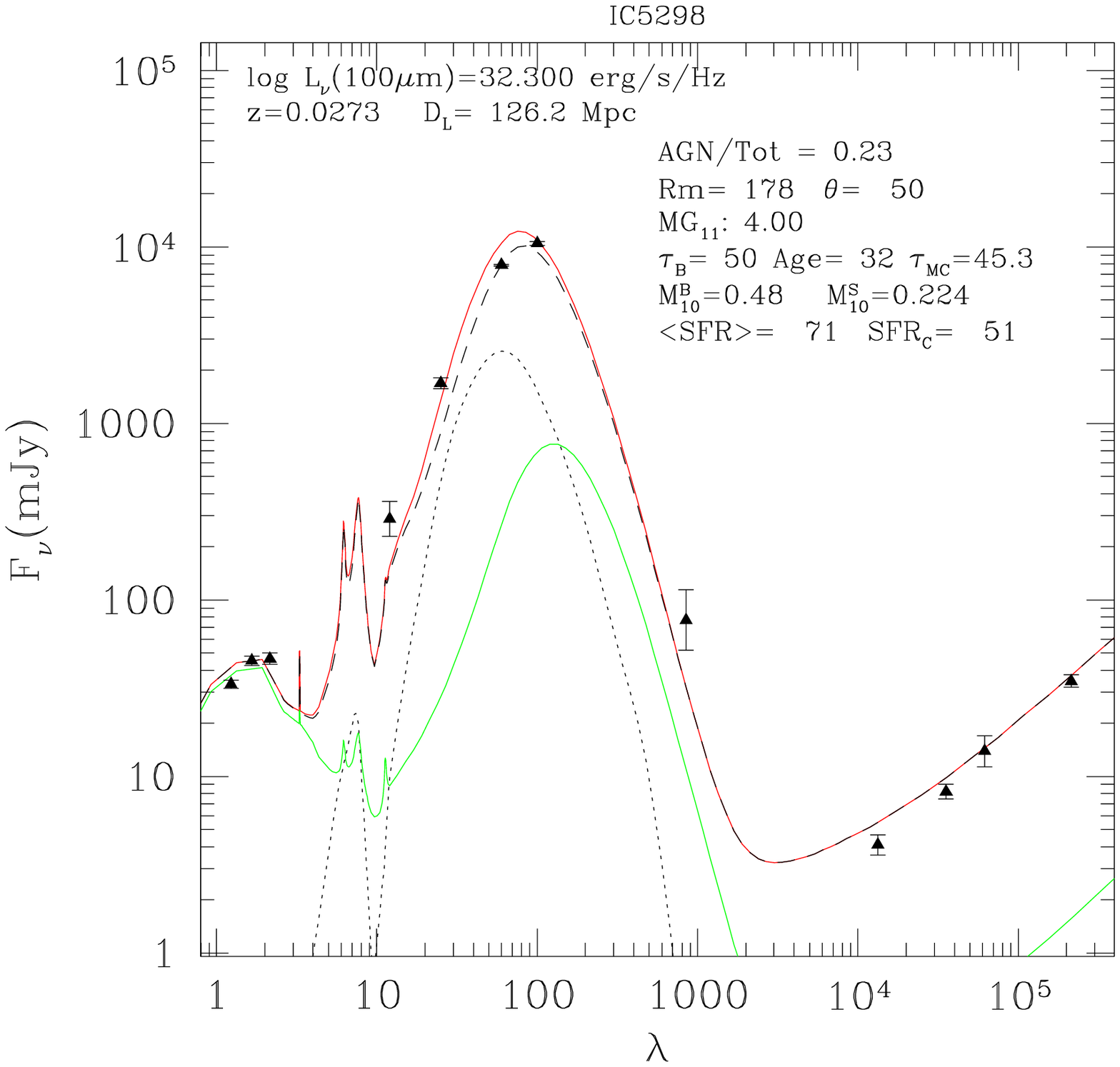}} &
\resizebox*{0.32\textwidth}{!}{\includegraphics{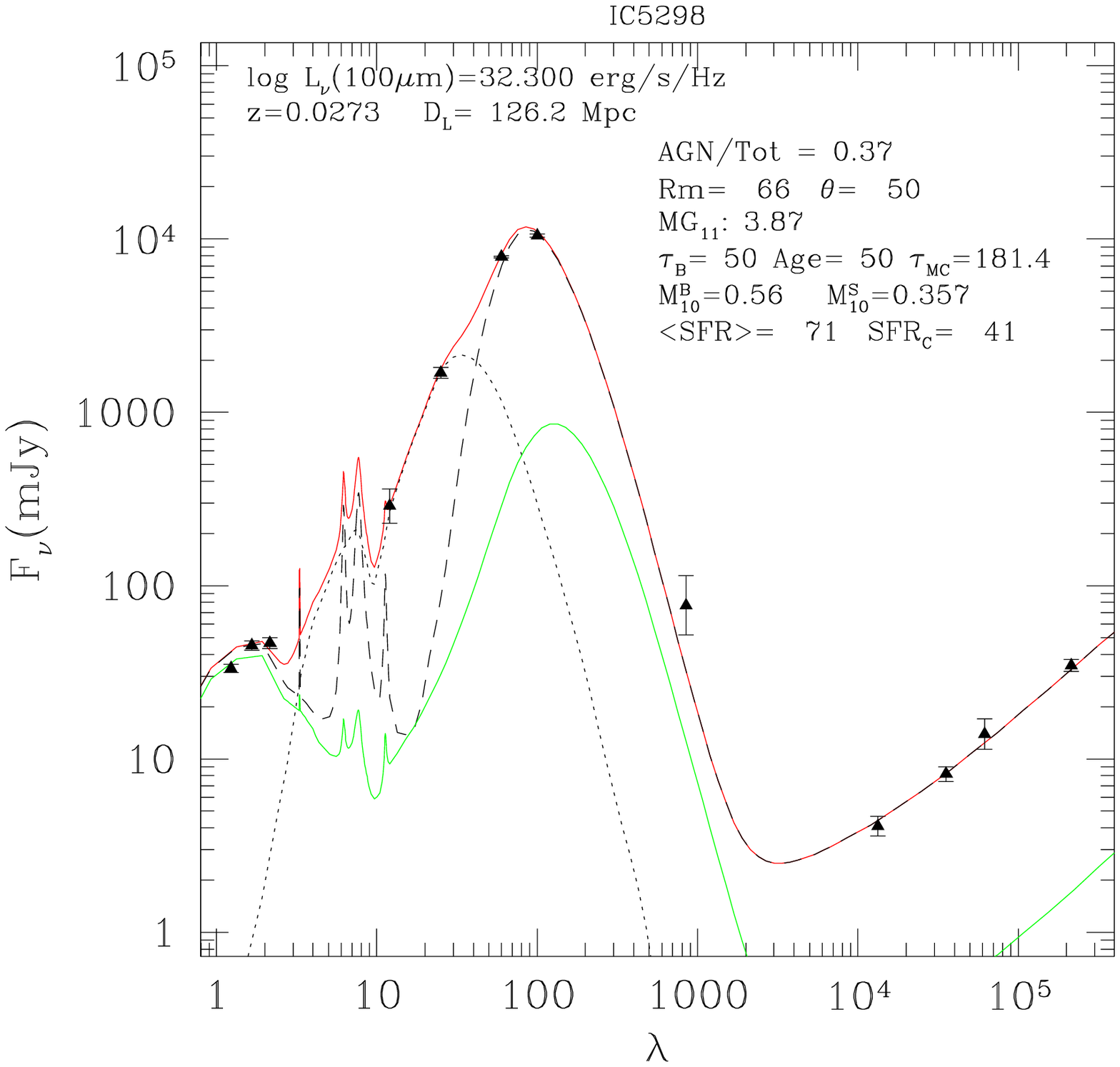}} 
\end{tabular}}
\caption{Model fits to observed SED of IC 5298.
A starburst  (dashed line) is combined with an AGN model (dotted line)
to obtain the total SED (upper solid line). The fractional contribution of the AGN to the
$8-1000\;\rm \mu m$ flux is indicated as AGN/Tot.  The lower solid line
shows the contribution of the underlying quiescent disk.} For other quantities
reported in the figure see text.  Left panel: without AGN component
and small optical depth in molecular clouds, $\tau_{\rm MC} = 15.1$ (a
gas/dust ratio of 300 is required in this model).  Middle panel: a fit
with the AGN component and $\tau_{\rm MC} = 45.3$.  Right panel: a fit
with the AGN component and large optical depth in molecular clouds.
Notice the large M$_{\rm Burst}$/M$_{\rm Gal}$ required by the model
with small $\tau_{\rm MC}$. 
\label{s1}
\end{figure*}

\subsection{Mapping \& Model-fitting}

Mapping and model-fitting were carried out in a homogeneous manner
 within {\sc difmap} (Shepherd 1997). Data were inverted onto a large map (1024 x 1024 pixels of 0.35 x 0.35
arcsec) the central quarter of which was cleaned (H{\"o}gbom 1974)
until the peak in the residual map $< 4
\sigma$ (where here $\sigma$ refers to the r.m.s. level of the residuals) and
 continued cleaning resulted in no further reduction in the residual
level. For the stronger sources the data were then self--calibrated
and re-cleaned until convergence. The data were re--mapped on to 512 x
512 pixels of side 0.35 arcsec and any further cleaning necessary was
performed in a clean window just big enough to contain the high
residuals surrounding the source position. When further cleaning
resulted in no improvement in the reduction of the residual level,
maps were restored.

Our maps show that, with the possible exception of NGC 7469, all
sources are unresolved at an angular resolution of approximately 2.5
arcsec, as expected from the results of Condon et al. (1991).
Flux densities were obtained directly from the restored maps  and from
modelling the sources in the visibility domain with {\sc difmap:
modelfit}.  An initial model for the source was taken from the map; this 
consisted of its flux, major and minor axes, position angle, the
radial distance and position angle of the source from the phase centre
and the form (e.g. point--like or Gaussian). {\sc difmap: modelfit}
transforms the image plane model components into the u--v domain to
enable a direct comparison of model visibilities with the observed,
calibrated visibility data.  A set of iterative adjustments were made
to each of the model parameters to minimize the model's
$\chi^2$--statistic. When successive iterations failed to provide an
improvement and $\chi^2_r< 1$ (where $\chi^2_r$ is the reduced
$\chi^2$--statistic) the iterations were halted. All fluxes presented
 in Table \ref{measuredfluxestable} are derived from source models with $\chi^2_r<0.85$.

\section{Results}\label{results}

\subsection{Radiometry}\label{radiometrysection}

The accuracy of our radiometry is critically dependent on the flux
densities we determined for the calibration sources.  In Table
\ref{calibratorfluxes} we list the assumed flux densities of the flux
calibrators (3C 48 and 3C 147) and the dependent measured fluxes of
the phase calibrators.  We note that, as expected, the phase
calibrator with the worst determination of flux, 0927+390 (with flux
density errors at the level of almost 1.5 percent), was that used on
January 3 for source IRAS 08572+39.

The error estimates were calculated based on the assumption that the
term due to measurement noise and that due to errors in gain
calibration (and therefore proportional to the measured flux) may be
added in quadrature. The first term was estimated by the noise in the
background of the uniformly weighted map, the second by the fractional
error in the associated phase calibrator's flux density (Table
\ref{calibratorfluxes}). The phase calibrators were chosen to be near
to the source and as opacity and gain curve corrections were performed
on loading data into {\sc aips}, no further dependence on zenith angle
was considered.
\begin{table} \caption{Parameters for a disk galaxy model. See
text for details.}
\begin{tabular}{cccccc}
\hline
\hline
$\nu$  & k & t$_{\rm inf}$ & x &
M$_{\rm Low}$ & M$_{\rm Up}$ \\
\hline 0.5 Gy$^{-1}$ & 1 & 9~Gy &1.35 & 0.15 M$_\odot$ &120
M$_\odot$ \\
\hline
\end{tabular}
\label{chem}
\end{table}
We expect flux densities measured directly from visibility data to be
more reliable than those determined from the map, being independent of
imaging errors and of cleaning.
 These measured source flux densities are presented in Table
\ref{sourcefluxes}. 
$S_{\rm uv}$ is the flux density determined from iteratively fitting
model visibilities to the calibrated data as described
above. Positions given in Table \ref{sourcefluxes} refer to the phase
centres and $r$ and $\phi$ to the radial offset and position angle
(north through east) of the peak from the phase centre, as determined
by the model--fitting procedure.

\begin{figure*}
\resizebox*{0.9\textwidth}{!}{\includegraphics{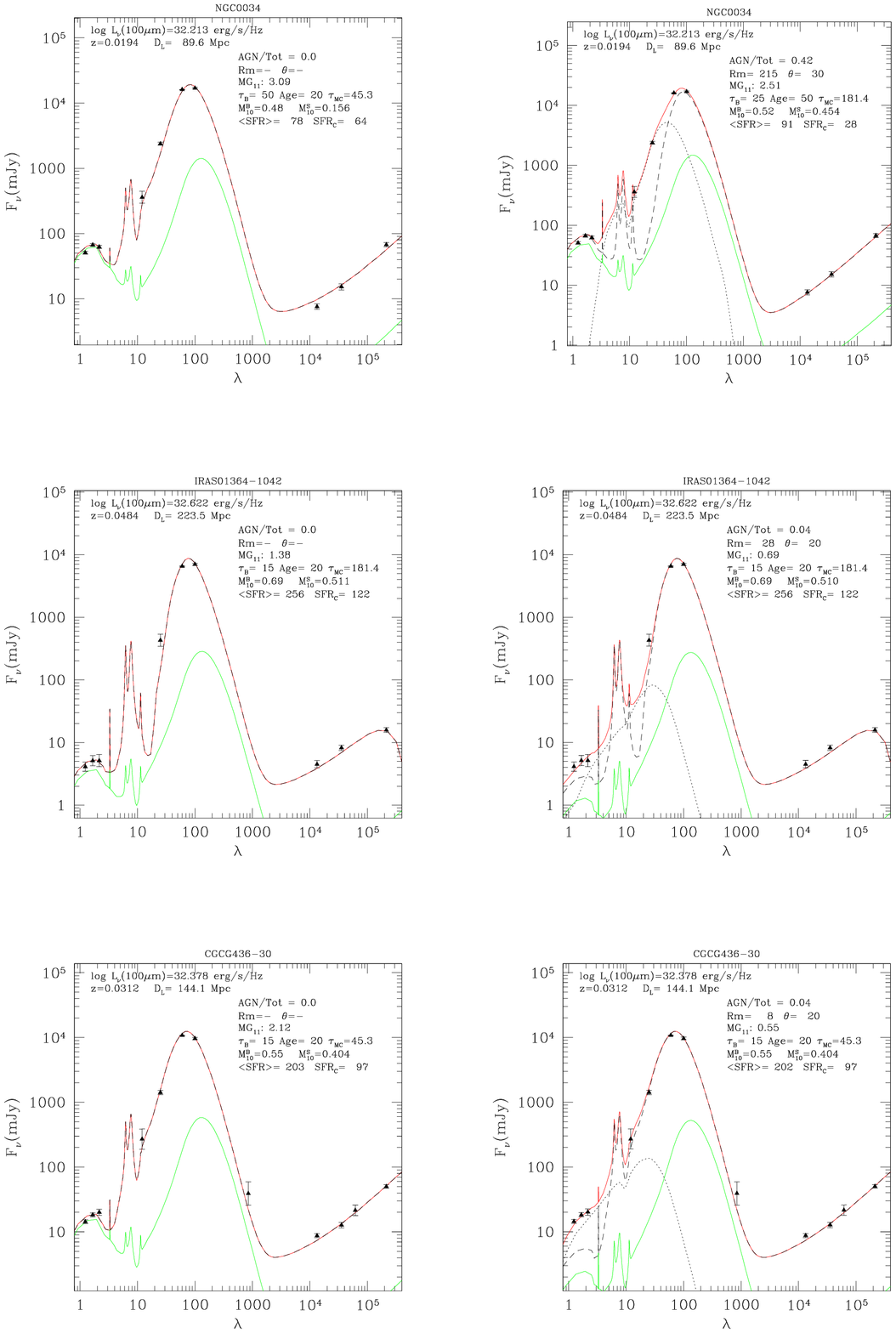}} 
\caption{Best fits of the SEDs of NGC 34,  IRAS 01364-1042 and CGCG 436-30
without (left panel) and with  AGN component (right panel). 
\label{s2}}
\end{figure*}

\begin{figure*}[t]
\resizebox*{0.8\textwidth}{!}{\includegraphics{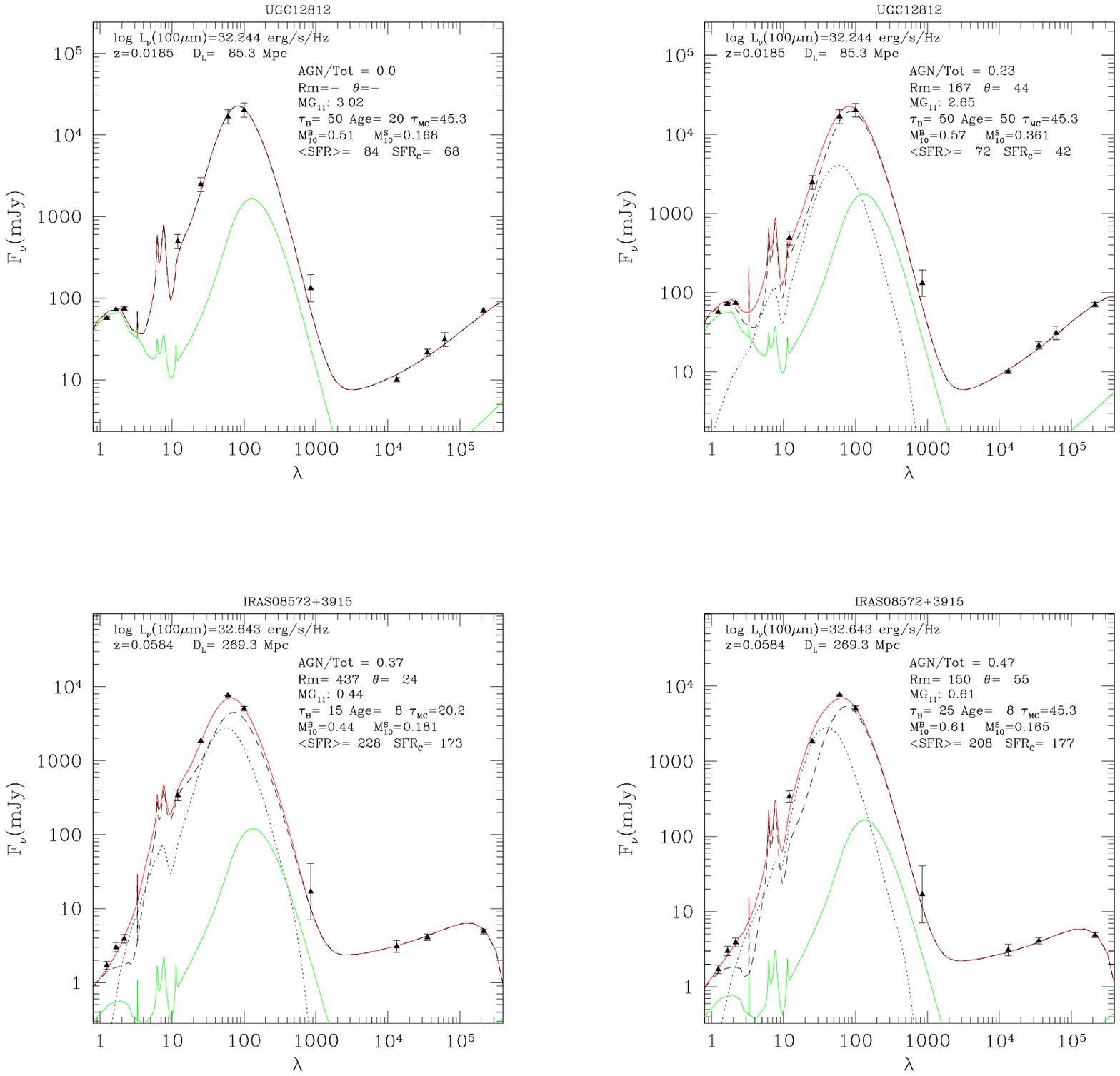}}
\caption{Fits of the SEDs of  UGC 12812 and IRAS 08572+3915.
In the case of IRAS 08572+3915 the presence of the AGN is evident from the
NIR fluxes and the fits with different  $\tau_{\rm MC}$ are shown.
\label{s3}}
\end{figure*}

\begin{figure*}
{\centering \begin{tabular}{ccc}
\resizebox*{0.32\textwidth}{!}{\includegraphics{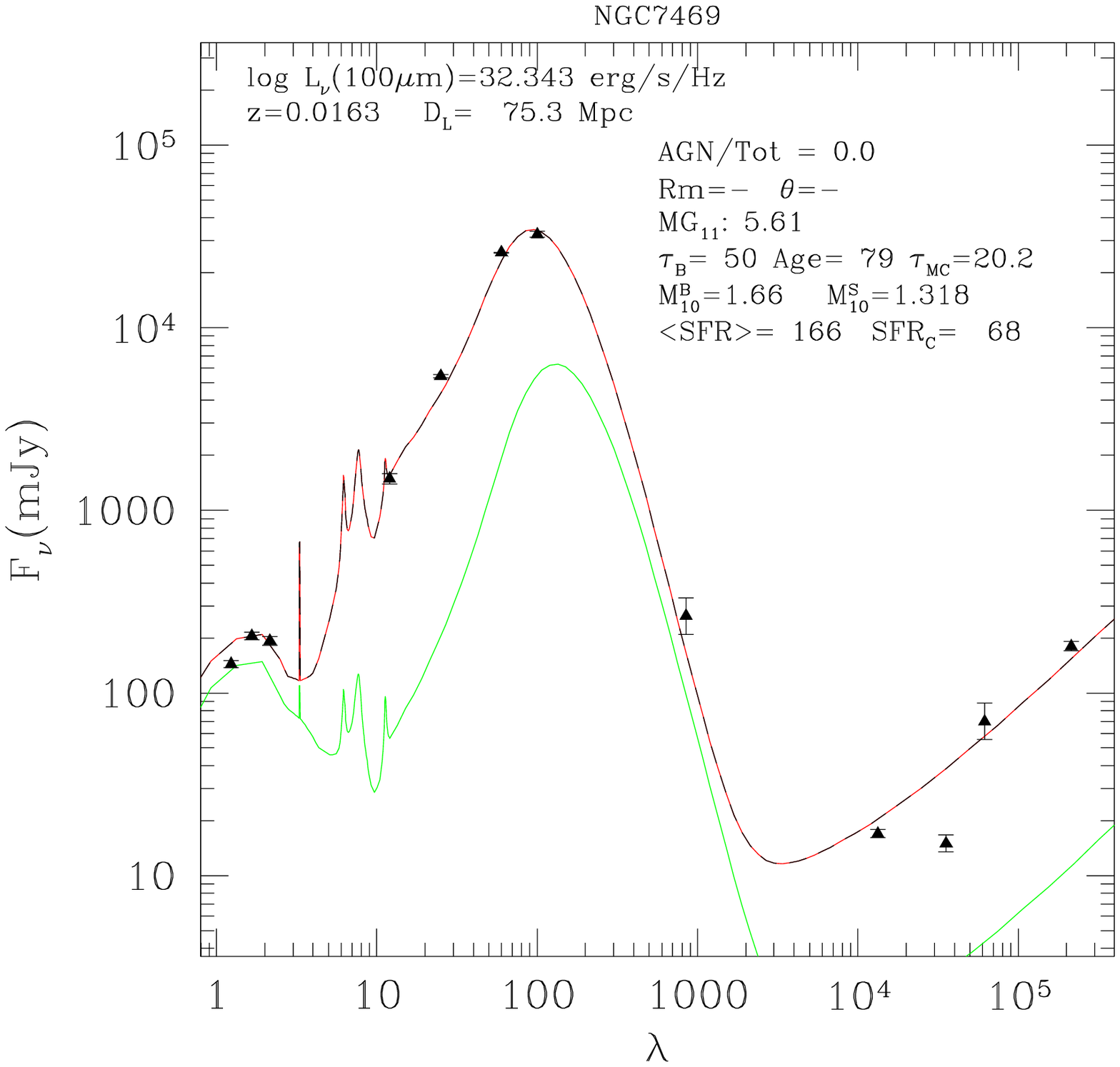}}  &
\resizebox*{0.32\textwidth}{!}{\includegraphics{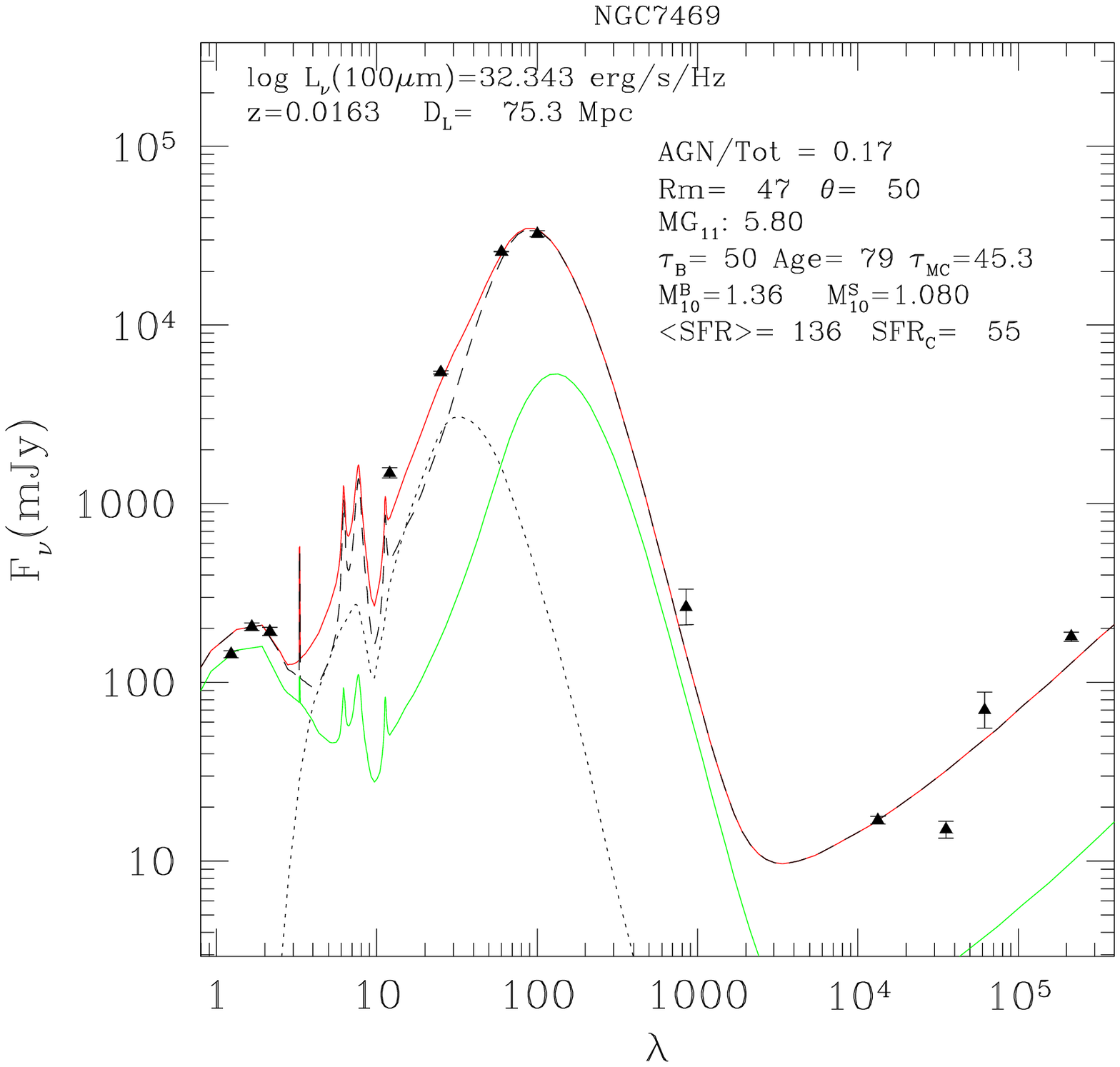}} &
\resizebox*{0.32\textwidth}{!}{\includegraphics{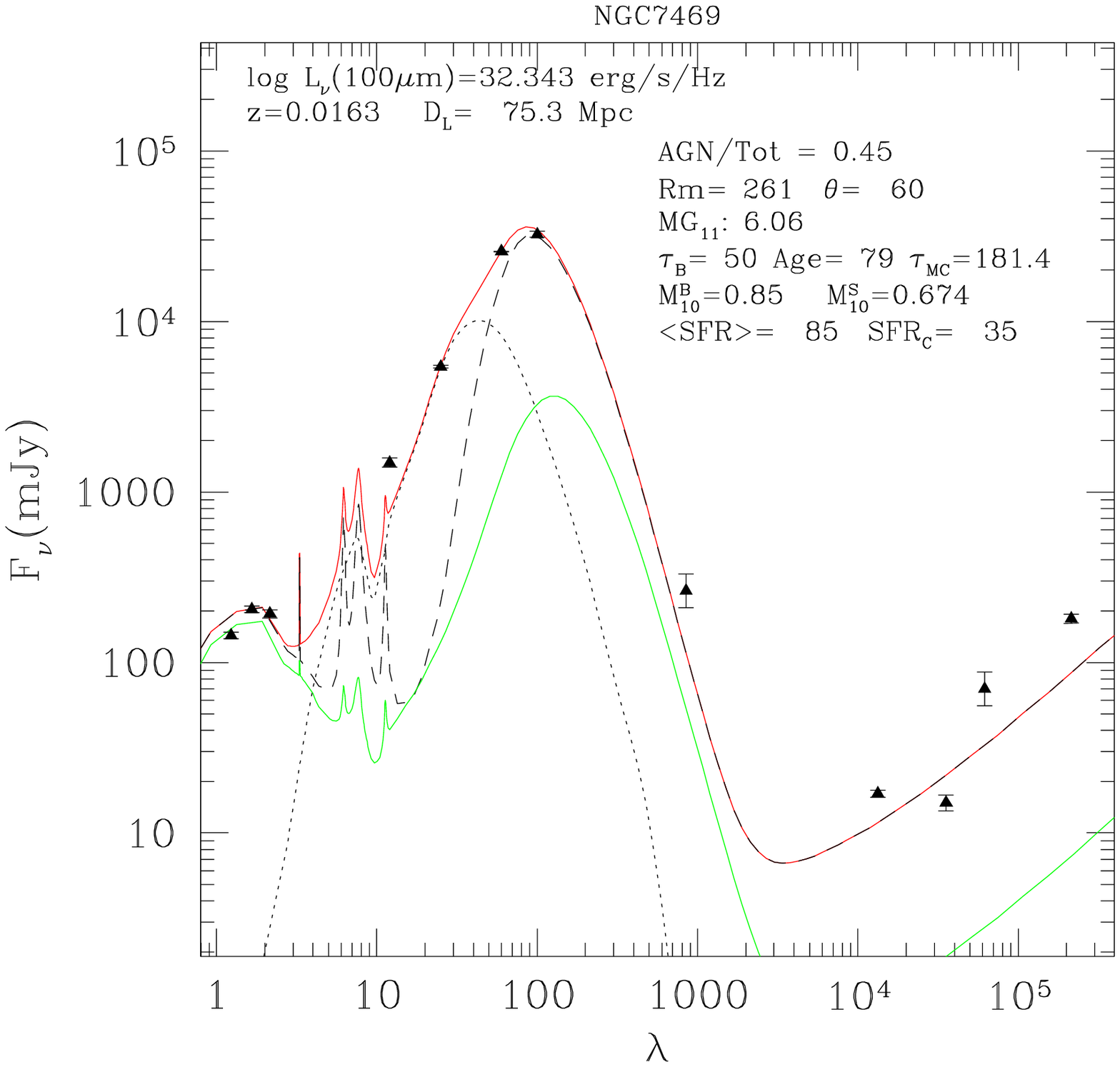}} 
\end{tabular}}
\caption{Fits of the SED of NGC 7469 at increasing AGN contribution.
Notice that consistency between FIR and Radio data exclude a large contribution
of the AGN in this galaxy.
\label{s4}}
\end{figure*}

\subsection{Comparing fluxes at different resolutions}

It is important to note that data at different frequencies were obtained from
observations at different resolutions 
 and that all the radio data are from observations at much higher resolution 
 than that of IRAS ($\sim 3$ arcmin at $100\;\rm \mu m$). For sources extended in the
radio this could bias flux measurements at higher
resolutions to lower values as flux becomes resolved out of
interferometric maps. For the IRAS fluxes we also need to consider the
potential of confusion with nearby sources.
Using the $1.4\;\rm GHz$ maps of Condon et al. (1990) we find that all of our
sources are smaller than $5$ arcsec 
 diameter in the radio
and often smaller than $2$ arcsec. 
 Our $23 \; \rm GHz$
observations will therefore not underestimate the fluxes in these
objects because of their angular sizes. The only data for which
resolution may be a problem are the VLA A-array observations at 8.4
GHz of Condon et al. (1991). These have a resolution of 0.25 arcsec
and for the source NGC 7469 some flux is almost certainly not
detected. The $8.4\;\rm GHz$ flux for this object is thus a lower limit, as
indicated in Table \ref{sourceproperties}.

We investigated the possibility of contamination of the IRAS fluxes by
using NVSS images to look for sources sufficiently close to our target
sources to cause confusion at  $100\;\rm \mu m$. 
 We then made
use of the FIR/radio correlation to estimate the proportion
of the IRAS flux contributed by the companion. The IRAS fluxes for 3
objects were thus estimated to have a small contribution (up to 10 percent) from a nearby source;
these are the only three from our sample that show evidence of being
resolved in the IRAS data of Soifer et al. (1989). The IRAS fluxes for
these sources were reduced as follows. NGC 7469: 7 percent; IC 5298: 10 percent;
UGC 12812: 3.5 percent. We note that these corrections are small and at
least for the 12 and $25\;\rm \mu m$ fluxes are of the same order as
the errors on the fluxes.

\section{Analysis of the spectral energy distribution}\label{sed}

Observations of the spectral energy distribution from the FIR to the radio have been
 complemented by flux measurements in the NIR J, H and K bands by the 
 Two Micron
 All Sky Survey (2MASS e.g. Jarrett et al., 2000). Berta et al., 2003 have
 convincingly shown that a J-K flux excess, over that
 expected from a pure starburst, is a strong indicator of the presence
 of an AGN.  Further signature of the presence of an AGN comes from
 the dust temperature derived from fitting IRAS fluxes
 (eg. Sanders et al., 1988b).  The contribution of an AGN in the radio
 is assumed to be zero (see Section ~\ref{discussion})  and we take
 the radio emission to be due entirely to star formation. We fit the
 spectral energy distribution with a combination of starburst models
 generated with the recent version of GRASIL (Silva et al. 1998, 
 Bressan, Silva \& Granato 2002), that includes radio
 emission from star forming regions, and  AGN models generated with
 the radiative transfer code developed by Granato \& Danese (1994).

\subsection{Starburst models}
The starburst model is described by the superposition of a burst of star
formation on a quiescent disk.  The SFR in the disk galaxy is obtained
by an infall chemical evolution model with a Schmidt law and a
Salpeter initial mass function (IMF) from 0.15 M$_\odot$ to 120
M$_\odot$.

The parameters of the chemical evolution models, i.e. the infall
time-scale (t$_{\rm inf}$) and the SF law, SFR=$\nu \, M_{\rm g}^{\rm
k}$ are shown in Table \ref{chem}.

At an epoch of 11.95 Gy the star formation rate is enhanced to mimic
the burst. The intensity of the SFR is fixed by assuming that the mass of stars
formed during the burst is  a fraction of the underlying disk mass 
and by the  e--folding time--scale taken to be $t_{\rm b}\;$= 10, 15, 25 or 50 My.

Half of the gas mass was assumed to be in the form of molecular clouds of
$10^6\;\rm M_\odot$, with the remaining gas distributed as a diffuse component.
Stars form at the centre of these molecular clouds and suffer an extinction
given by the cloud optical depth, $\tau_{\rm MC}$. The obscuration time, t
$_{\rm esc}$, was set to linearly decrease with time from $t_b$ down to a
minimum of 3 My (characteristic of normal galaxies). Older stars are attenuated
only by the diffuse component. The dust/gas ratio was selected to be 0.01. PAH
features and temperature fluctuations of small grains were taken into account
as described in Silva et al. (1998).  Our main conclusions are not affected by
the details of these parameters.  However, since the mid
infrared region (12 to $25\; \rm
\mu m$) is sensitive to the optical depth of the molecular clouds
($\tau_{1 \rm \mu m}^{\rm Mc}$), we considered three different
values of $\tau_{1 \rm \mu m}^{\rm Mc} =\;$20, 45 and 180.

The ensemble of older stars, molecular clouds and diffuse component
of the burst follow a spheroidal geometrical distribution (see Silva et al 1998
for more details).

\subsection{AGN models}

Dusty AGN SEDs were selected from libraries generated with the
model of Granato \& Danese (1994) and Granato, Danese \&
Franceschini (1997). Each library consists of several hundred cases, belonging to a given geometrical class, in which typically
4 to 5 parameters are varied, assigning 3 to 4 widely differing values to each.
 Libraries comprise both anisotropic flared
discs and tapered discs (for definitions see for instance
Efstathiou \& Rowan-Robinson 1995). The effects of various
parameters on the predicted SED have been explored in Granato \&
Danese (1994) and Granato et al.\ (1997). Also included in the
libraries are  models in which the size distribution of grains
extend to radii larger than the standard value, as suggested for example by Maiolino et al. (2001a, b) and Galliano et al.
(2003). We refer the reader to Galliano et al. (2003, in press) for a full description of these libraries.

\subsection{Model Fitting}

Model fits to the SEDs described above are shown in Figures \ref{s1} to
\ref{s4}; their parameters are shown in Table \ref{masses}. 
Here 
$M_{\rm burst}$ is the total mass of gas and stars in units of
10$^{10}\; \rm M_\odot$;
 $\tau_{\rm b}$ is the e-folding time of the SFR in units of 10$^{6}\;$y;
 $age$ is the current age of the burst in units of 10$^{6}\;$y;
 $M_\star$ is the current mass in stars in units of 10$^{10}\;$M$_\odot$;
 $\overline{SFR}$ is the average SFR in units of M$_\odot$/y over the lifetime of the
 burst;
 L$_{\rm AGN}$/L$_{\rm Tot}$ is the fractional  contribution of the AGN to the total
luminosity from 8 to $1000\; \rm \mu m$;
 $\theta$ and Rm are the inclination and the ratio of the external to internal radius of the
torus, respectively. 
$M_{\rm Gal}$ is the mass of the
old disk, in units of  10$^{11}\; \rm M_\odot$, as determined by the
fit to the NIR fluxes. It is worth noting that, when an AGN is
present, and in particular when this is indicated by  the shape of the
SED in the NIR, $M_{\rm Gal}$ must be considered an upper limit to
the mass of the galaxy. This is because the AGN contributes flux in excess of
that from the old stellar disk at NIR wavelengths.

$SFR_{\rm fir}$ is the current star formation rate derived from the $L_{\rm fir}$ of the modelled starburst
spectrum. This is added as a comparison showing how the SFR might be
estimated using one of the standard measures (Kennicutt, 1998).
 Kennicutt's (1998) calibration was obtained from the ratio of the
bolometric luminosity to SFR in a model of constant star formation
over the last 100 My. Since it refers to a Salpeter IMF between
$0.1\;\rm M_\odot$ and $100\;M \rm _\odot$ we have divided the values
obtained with the Kennicutt relation by 1.16 to account for our
different limits on the IMF (between $0.15\; \rm M_\odot$
and $120\;\rm M_\odot$).  

Finally the last column reports the value of the merit function of the
fit. The merit function is calculated as
\begin{equation}
{\sl F}=\frac{1}{N}\Sigma \frac{(M_i - O_i)^2}{E_i^2}
\end{equation}
where M$_i$, O$_i$ and E$_i$ are the model values, the observed values and
the observational errors respectively, N is the number of passbands used for
the fit.

The models shown in Figures \ref{s1} to \ref{s4} 
result from the combined contribution (solid line)  of a
starburst model (dashed line) and an AGN model (dotted line).  In the
plots we also show the instantaneous, current star formation rate
$SFR_{\rm c}$.

The starburst model includes the contribution of the
underlying quiescent star forming disk. The spectral energy
distribution of this disk is shown as the lower solid line. Its
contribution to the total FIR and radio emission is negligible since
its star formation rate amounts to only a few M$_\odot$/yr while that
of the starburst is, in general, an order of magnitude
larger. However, due to the cooler temperature of the dust that
characterizes the quiescent disk, its contribution to the
sub-millimetre SED may be significant. This should serve as a warning against
fitting the sub-millimetre SED of infrared sources with a single
component starburst model. In the near--IR the old population of the
disk always provides  a significant contribution to the SED of the
galaxy, in agreement with the results of Mayya et al. (2004).

\section{Discussion}\label{discussion}

Almost since the discovery of the far-infrared/radio correlation
it has been suspected that the scatter in the relation has a contribution from the
delay between the infrared and radio emission during the star
formation process (e.g. Helou, Soifer \& Rowan-Robinson, 1985,
Wunderlich \& Klein, 1988). Bressan, Silva \& Granato (2002)
investigated this idea with a model in which the scatter in the value
of $q$, the FIR/radio ratio, in ULIRGs is caused by evolution i.e. the
effect of an intense, short duration burst that is quenched
rapidly. In particular,  high values of $q$, relative to the average
for star forming galaxies, should be typical of young bursts because
of the delay between FIR emission and radio emission. Moreover, a high
value of $q$ should be  associated with flat radio slopes because the
radio emission in the early phases  of the burst is dominated by
free-free emission. This suggestion is generally confirmed by the
present investigation.  However, it is clear that this simple picture
is complicated by the possible presence of AGN, which may contribute a
significant fraction of the FIR luminosity shortward of $60\;\rm \mu
m$.

Above  $60\;\rm \mu m$, the FIR to radio SED provides fairly good
constraints on the starburst component. In particular, flat radio
slopes are associated with an excess of FIR flux with respect to the
radio emission, in agreement with the predictions for young starburst
models.

In fitting the model to the data we have neglected the possible
contribution of an AGN at radio wavelengths.  It has been shown, using
VLBI observations of ULIRGs with compact radio cores (Smith et al.,
1998), that the average value  at $1.6\; \rm GHz$ of $S_{\rm
VLBI}/S_{\rm Total}= 0.12$. These observations suggest that, in our 
sample,  the  AGN contribution in the radio is small.  

Furthermore, the shape of the radio spectrum, 
both in objects with and without the signature of AGN contamination
in the mid-IR, can be reproduced very well by starburst models.

If the contribution of the AGN were significant at radio frequencies the
relative contributions of the starburst and AGN in the models would require
fine tuning in order to reproduce the observed radio spectrum.
In particular, it would be difficult to explain
the variation of the radio slope as one moves from the  
synchrotron dominated (1.4GHz) to the free-free dominated (23GHz)
regime.

It is also evident from our fits that the starburst almost entirely dominates
the SED at $100\;\rm \mu m$, even when there is a strong contribution from the
AGN to the total FIR luminosity.

The ratio between $100\;\rm \mu m$  and radio fluxes  is
always well reproduced by starburst models. The presence of a
radio--loud AGN  will affect this ratio more than that between the 
total FIR and radio fluxes.

Thus, we argue that the contribution of the AGN at radio wavelengths must be
small in our objects, $\leq$ 10\%. This does not significantly modify our
conclusions concerning the star formation history of the galaxy, if one
considers that starburst radio fluxes change by about one order of magnitude
from 1.4GHz to 23GHz.

Whether this is a common characteristic of 
the compact ULIRGs remains to be investigated by extending our method 
to the whole sample of Condon et al. (1991).

Observations close to $23\; \rm GHz$ are decisive because 
they significantly extend the frequency range that is  
not affected by free-free absorption. Condon et al. (1991) have
argued that
$1.4\; \rm GHz$  radio fluxes in compact ULIRGs may be 
affected by free-free absorption. Furthermore, 
Bressan, Silva \& Granato (2002) have shown that this effect may be mistaken
for evolution, i.e. a shallower radio slope determined from the $1.4\; \rm GHz$
radio flux may result either from free-free absorption or a younger
sturburst.

Between $1.4\; \rm GHz$ and $23\; \rm GHz$ 
synchrotron emission
fades by about an order of magnitude,  disclosing
the thermal emission from 
ionized gas, which is a more direct tracer of the ongoing star formation.

Since thermal and non-thermal radio emission appear in
proportions that are  dictated by the recent history of star formation,
observations of the SED over such a wide range of frequencies prove
to be essential for the determination of the burst parameters.

\begin{figure}
{\centering 
\resizebox*{0.45\textwidth}{!}{\includegraphics{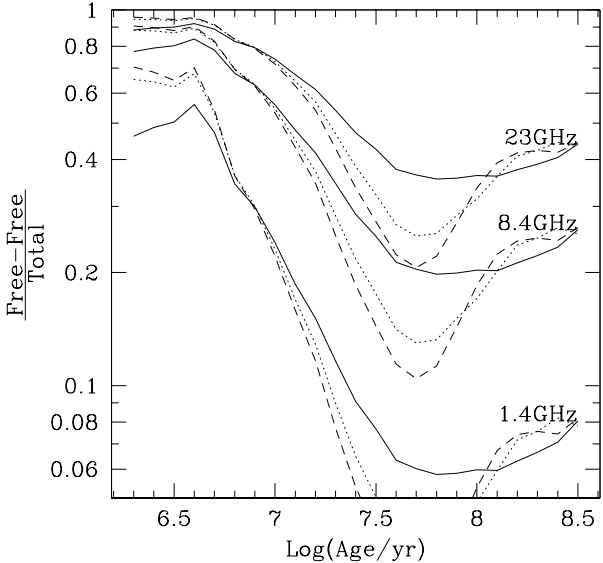}} 
}
\caption{Fractional contribution of free-free emission to 
total radio emission as a function of time in different starburst models.
The SFR is
 declining exponentially with a characteristic time $\tau_{\rm b}$=50Myr
(solid line), $\tau_{\rm b}$=25Myr (dotted line) and $\tau_{\rm b}$=15Myr (dashed line).
}
\label{frefre}
\end{figure}

To clarify this point, figure \ref{frefre} depicts the fractional
contribution of free-free emission to the total radio emission for
different frequencies as a function of the burst parameters: age and
SFR decay--time in our models. 
Free-free emission from
ionized HII regions always provides a significant contribution to the
$23\; \rm GHz$ flux, being less than 50 percent only during the  late evolution
of the burst, when the radio emission is dominated by supernovae from
less massive stars.  At lower frequencies not only does the
fractional contribution decrease but during the early phases of the
burst it does not provide 100 percent of the radio emission, as may be
expected,  due to the supernova emission of the old stellar disk.

For bursts younger than 20 My more than half of the $23\; \rm GHz$
flux is contributed by free-free emission from HII regions, while at
ages $t < $ 10 My this fraction rises to 80 percent. Estimates of the
ionizing photon flux (and therefore the SFR) based on the free-free
emission may be affected by uncertainties  in the fraction of ionizing
photons lost by internal dust or  escape (which can be  as large as 30
percent, Panuzzo et al., 2003)  but they are independent of the
uncertainties that affect synchrotron emission. Effects such as the
strength of the magnetic field and inverse Compton cooling on the
strong radiation field of the starburst are discussed in Bressan,
Silva \& Granato (2002).

In three sources,  NGC0034, IRAS01364-1042 and IRAS08572+3915 
the determination of the burst parameters would not be possible
without high frequency observations. Moreover in IRAS01364-1042  and  IRAS08572+3915 
the $23\; \rm GHz$ data allow us to conclude that the $1.4\; \rm GHz$ 
fluxes
are affected by significant free-free absorption.
In all other galaxies the $23\; \rm GHz$ observation 
unambiguously fixes the thermal contribution and then the
age of the starburst.

Thus, the combination of 
accurate radio measurements with FIR data 
provides a good estimate of the relative contributions  of the AGN and
the starburst.

In some cases, the presence of an AGN is revealed  by the NIR colours
 showing the shape of a typical, dominantly power--law spectrum,
instead of the knee--dominated starburst spectrum. However,
by means of these colours alone it is not possible  to constrain 
the fractional contribution of the AGN, because the colours
are a function of the torus model (optical depth, inclination etc.)

In summary, we find that all the compact 
ULIRGs in our sample are associated with
strong,  young bursts of star formation. The SF rates, averaged over
the age of the burst, exceed $100\;\rm M_\odot$/y. Typical total gas
masses involved in the bursts are around 10$^{10}\;\rm M_\odot$ and
the mass of stars formed so far is between 10 and 40 percent of this
mass.

\subsection{High optical depths}

Our models generally require a large  molecular cloud
optical depth at 1~$\mu$m, typically
$\tau_{1 \rm \mu m} \simeq 50$, but values as high as $\sim 180$ are necessary for
IRAS 01364-1042. One might ask whether or not such high values are reasonable.
Remembering that the extinction  within our molecular clouds has a 
foreground screen geometry, an
optical depth of 50 corresponds to an extinction of 54 mag. Assuming the Galactic
interstellar extinction curve (Whittet, 1992) this implies a visual extinction
$A_{\rm v} \simeq 130\;\rm mag$. A corresponding total neutral ISM column density
can be derived from the empirical relation of Bohlin et al. (1978) for sight-lines
to reddened stars in the Galaxy. Again assuming the Galactic extinction curve we
find
$N_{\rm H} = 1.86 \times 10^{21} \, A_{\rm v} = 2.5 \times 10^{23}\;\rm
atoms\;cm^{-2}$ where $N_{\rm H} = \rm N(HI + 2H_{2})$ is the total neutral ISM
column density in atomic and molecular gas. Several studies have measured the
molecular gas column densities towards the centres of luminous and ultraluminous
infrared sources using interferometric CO data (e.g. Downes \& Solomon, 1998;
Scoville et al., 1989; Sanders et al., 1988a). Peak column densities of
$N_{\rm H_2}\sim 10^{23}\;\rm cm^{-2}$ are commonly found. The column density of
the atomic gas has also been measured for those sources that show HI absorption 
(e.g., Mundell, Ferruit \& Pedlar, 2001; Beswick et al., 2003; Clemens \&
Alexander, 2003) and for an assumed spin temperature of $100\;\rm K$, peak column
densities of $N_{\rm HI}\simeq 10^{22}- 10^{23}\;\rm cm^{-2}$ are found. The
column density of gas implied by our extinction measurements are therefore quite
consistent with measured values for similar, extreme objects. In addition to these
measurements of the gaseous ISM, which indirectly constrain the extinction, there
have recently been direct measurements which make use of mid-infrared spectral
features. Haas et al. (2001) find an extinction of $A_{\rm v} \sim 110\;\rm mag$
for Arp 220 by comparing the $7.7\;\rm \mu m$ PAH feature with the
$850\;\rm \mu m$ continuum flux and Soifer et al. (2002) find values of
$A_{\rm v} = 80 - 150\;\rm mag$ using spectroscopy of the $9.7\;\rm \mu m$
absorption line. Values of $\tau_{1 \rm \mu m} = 50$ are therefore quite
reasonable and values 3 times higher should not be excluded.

\begin{table*} \caption{Star-burst and AGN properties of our
compact luminous infrared sources.}
\begin{tabular}{lrrrrrrccrrrrr}
\hline
\hline
Name              & $M_{\rm b}^{(a)}$  &  $\tau_b^{(b)}$ &$age^{(b)}$  & 
$M_\star^{(c)}$& $\overline{SFR}^{(d)}$ & $\tau_{\rm MC}$ &
L$_{\rm AGN}$/L$_{\rm Tot}^{(e)}$  & $\theta$ & Rm & $M_{\rm Gal}^{(f)}$ &
$SFR_{\rm fir}^{(g)}$ & {\sl F}\\
\hline
NGC0034           &  0.48   & 50     &  20   &  0.156  & 78  &  45.3   & -    & -  &-    &  3.08 & 47   &  29.3 \\
NGC0034           &  0.52   & 25     &  50   &  0.454  & 91  & 181.4   & 0.42 & 30 & 215 &  2.51 & 30   &  6.19  \\
\\
CGCG436-30        &  0.55   & 15     &  20   &  0.404  & 203 &   45.3  & -    & -  & -   &  1.89 & 87   &  83.2  \\
CGCG436-30        &  0.55   & 15     &  20   &  0.404  & 202 &   45.3  & 0.04 & 20 &8.5  &  0.55 & 87   &  5.57  \\
\\
IRAS01364-1042    &  0.69   & 15     &  20   &  0.511  & 256 &   181.4 & -    & -  & -   &  1.38 & 110  &  13.2 \\
IRAS01364-1042    &  0.69   & 15     &  20   &  0.510  & 256 &   181.4 & 0.04 & 20 & 27.5&  0.69 & 110  &  4.27  \\
\\
IRAS08572+3915    &  0.44   & 15     &  8    &  0.181  & 228 &  20.2   & 0.37 & 24 & 436 &  0.44 & 145  & 6.97 \\
IRAS08572+3915    &  0.61   & 25     &  8    &  0.165  & 208 &  45.3   & 0.47 & 55 & 150 &  0.61 & 127  & 26.8   \\
\\
NGC7469           &  1.66   & 50     &  79   &  1.318  & 166 &   20.2  & -    & -  & -   &  5.61 & 65   & 31.4   \\
NGC7469           &  1.36   & 50     &  79   &  1.080  & 136 &   45.3  & 0.17 & 50 & 47  &  5.80 & 53   & 53.2     \\
NGC7469           &  0.85   & 50     &  79   &  0.674  & 85  &   181.4 & 0.45 & 60 & 260 &  6.05 & 38   & 103.0    \\
\\
IC5298$^h$        &  5.13   & 15     &  79   &  5.109  & 643 &  15.1   & -    & -  & -   &  1.96 & 64   & 12.1   \\
IC5298            &  0.48   & 50     &  32   &  0.224  & 71  &  45.3   & 0.23 & 50 &178  &  3.99 & 48   & 29.4   \\
IC5298            &  0.56   & 50     &  50   &  0.357  & 71  &  181.4  & 0.36 & 50 &66   &  3.86 & 40   & 1.96   \\
\\
UGC12812          &  0.51   & 50     &  20   &  0.168  & 84  &   45.3  & -    &-   &-    &  3.02 & 50   &  7.52 \\
UGC12812          &  0.57   & 50     &  20   &  0.361  & 72  &   45.3  & 0.23 & 44 & 167 &  2.64 & 41   &  2.51 \\
\\
\hline
\end{tabular}
{\footnotesize\\ 
$^{(a)}$ Total mass of gas \& stars in units of 10$^{10}\;$M$_\odot$;   $^{(b)}$ time in  10
$^{6}\;$y;  $^{(c)}$ Current mass in stars in units of $10^{10}\;\rm M_\odot$;   $^{(d)}$ Average 
SFR in
units of M$_\odot$/y;  $^{(e)}$ Luminosity integrated from 8$\mu$m to $1000\; \rm \mu m$  $^{(f)}$
Total mass of the galaxy in units of $10^{11}\; \rm M_\odot$;  $^{(g)}$ SFR calculated from the
$L_{\rm fir}$ derived from the modelled starburst component;  $^{(h)}$ gas/dust ratio = 300
  
}

\label{masses}
\end{table*}

\subsection{Notes on Individual Sources}

{\bf NGC 34} (MRK 0938):  The presence of two nuclei separated by
approximately 1.2 arcsec in the mid-infrared (Miles et al., 1996) and
optical tidal tails indicate this galaxy is undergoing a merger
(e.g. Mulchaey, Wilson \& Tsvetanov 1996). The nature of the activity 
has been controversial, some authors
(e.g. V\'eron--Cetty \& V\'eron 1991) claiming a Seyfert 2 nucleus,
while others suggest a composite spectrum.  Mulchaey, Wilson \&
Tsvetanov's (1996) emission--line images show the galaxy to be a weak
emitter of [OIII]$\lambda 5007$ when compared with Seyfert galaxies in
their sample. In addition they argue its strong $\rm{H}\alpha$
emission, distributed over the entire galaxy, as indicative of a
starburst.

A fit with no AGN component is marginally consistent with the  radio
data. On the other hand, a large ($\simeq$ 40 percent) AGN contribution to
the $8-1000\;\rm \mu m$ luminosity is obtained by  assuming $\tau_{\rm
MC}=181$. In this case the AGN is quite dust enshrouded and the
spectrum falls off rapidly in the NIR. It is interesting to note that,
in both cases, the average SFR during the burst is of the order of
$80-90\;\rm M_\odot$/y.

{\bf CGCG 436-30:}  This galaxy is almost certainly interacting and is
detected in HI in both emission and absorption; the deduced  atomic
gas mass is greater than $3.4\times 10^9\;\rm M_\odot$ and must be
concentrated near the core in order to explain the absorption feature
(Mirabel \& Sanders 1988).

There is evidence of a compact radio core that may be difficult to
explain as an ensemble of young supernovae associated with the
starburst; the major contribution to the FIR and radio emission
originates on larger spatial scales (Lonsdale, Smith, \& Lonsdale,
1993;  Smith, Lonsdale, \& Lonsdale 1998).

The monotonically rising NIR fluxes require a small AGN
contribution but all measured fluxes longward of $12\;\rm \mu m$ can be
fitted with a pure starburst.

Our best fit SEDs, with or without the AGN component, indicate a
gas consumption time--scale of about 15 My and a starburst age of 20
My. The total mass  involved in the burst is $\simeq 5.5 \times
10^9\; \rm M_\odot$ and the residual gas mass $\simeq 1.5 \times
10^9\; \rm M_\odot$. The average SFR $\simeq 200\; \rm M_\odot$/y.

An upper limit to the mass of the galaxy $\simeq 1.9 \times 10^{11}\;
\rm M_\odot$ is provided by the fit of the NIR fluxes in the model
without the AGN component.

{\bf IRAS 0136-1042:}  The relatively low flux at $25\;\rm \mu m$
requires a large value of $\tau_{\rm MC}$ (181 in Fig. \ref{s2}). As
for CGCG 436-30, the NIR could require a small (4 percent) AGN
contribution, but the overall fit is also compatible with that of a
pure starburst.

An upper limit to the mass of the galaxy $\simeq 1.4 \times 10^{11}\;
\rm M_\odot$ is provided by the fit of the NIR fluxes in the model
without the AGN component.

{\bf IRAS 0857+3915:}   Consists of two nuclei separated by
$5$ arcsec. It is the north-west component which dominates in 
 both the infrared and radio (Sanders et al., 1988b; Condon et al., 1991).
Surace \& Sanders (2000) found evidence for stars no older than  100
My and despite the warm IRAS colours Veilleux, Sanders \& Kim (1999)
found no clear evidence for the presence of an AGN from NIR spectroscopy.

We fail to obtain a fit to the radio--100$\; \rm \mu m$ and 12-60$\;
\rm \mu m$ data simultaneously with a starburst model alone. As
expected from the J, H and K fluxes a significant AGN component is
required. Radio data themselves are well fitted by a starburst model, 
if one assumes $\tau_{ff}$$\simeq$0.7 at 1.4 GHz  (ie. significant
free-free absorption).  The AGN component is large  and  L$^{\rm
AGN}_{8-1000\; \rm \mu m}$/L$^{\rm Tot}_{8-1000\; \rm \mu m}$ (\rm
$L_{\rm AGN}/L_{\rm Tot}$)  ranges from 37 to 47 percent, in the case of a small or a  large
$\tau_{\rm MC}$, respectively.

The total mass  involved in the burst  $\simeq 5 \times 10^9
M_\odot$ and the residual gas mass  $\simeq 3.3 \times 10^9 \rm
M_\odot$.  The average star formation rate $\simeq 200 \rm
M_{\odot}$/y and  the recent star formation history is essentially
fixed by the shape of the SED from the radio to $100\; \rm \mu m$.

IRAS 0857+3915 is the only source common to both our study and the more extended sample 
of ULIRGs analysed by Farrah et al. (2003).
Our  estimate of the  AGN contribution to the total
FIR luminosity (Table \ref{masses}), is in very good agreement with their
quoted value ($\simeq$ 34\%), despite  their analysis being based on 
entirely different models (which make no use of the radio measurements).
 Using radio data we are able to provide a more accurate quantification of the
starburst parameters. 
Our derived age is 8 Myr versus Farrah et al.'s  upper limit of 57 Myr and our average
SFR is 228 M$_\odot$/yr against their quoted value of 173 M$_\odot$/yr. 
(Distance effects and IMF effects almost cancel out since
our luminosity is 17\% higher and our IMF limits
imply a 16\% lower total mass). As anticipated,
the $23\; \rm GHz$ observation allows the determination
of significant free-free
absorption at $1.4\; \rm GHz$, an important 
source of scatter around the FIR/Radio
correlation, not included in the analysis of Farrah et al. (2003).

{\bf NGC 7469:}   This source is thought to have a Sy 1.2 nucleus and
a circumnuclear starburst. Using high resolution mid-infrared
observations, Soifer et al. (2003) argued that the central source is
 an AGN rather than a nuclear starburst because of its very high
$12\;\rm \mu m$ surface brightness. Miles, Houck \& Hayward (1994) and
Nikolic et al. (2003) both found that PAH features in the mid-infrared
spectrum, thought to trace star formation, are found in the ring of
emission surrounding the nucleus but not in the nucleus itself. If
the hard ionizing photons from the AGN destroy PAH molecules this
supports the picture of a central AGN surrounded by a starburst ring.

Our best fit model has $L_{\rm AGN}/L_{\rm Tot} \leq 0.17$. At
mid-infrared wavelengths we predict the AGN and starburst
contributions to both be significant. This is the only source in our
sample for which spatially resolved mid-infrared data are available,
and in this case our model prediction is consistent with these data. A
larger contribution from the AGN is inconsistent with the mid-infrared
observations mentioned above, which show that significant flux comes
from both the starburst ring and the AGN. A larger AGN fraction is
also not consistent with the simultaneous fit of the FIR and radio
data.

{\bf IC 5298:}  Our best fit model has $L_{\rm AGN}/L_{\rm Tot}=0.36$,
a very high optical depth for the molecular clouds, $\tau_{\rm
MC}=181$ and an average SFR of $71\; \rm M_\odot$/y.

Wu et al. (1998a,b) found that IC 5298 has properties that are
intermediate between H II regions and liners; it was classified as a
Seyfert 2 by Veilleux et al. (1995) although the [O I] lines are
rather weak. Poggianti \& Wu (2000) classify this galaxy as e(a) type,
namely with H$\alpha$ in emission with moderate  equivalent with
($\simeq$52\AA) and H$\delta$ in absorption ($\simeq$4\AA). They
derived a SFR of about $92\; {\rm M}_{\odot}$/y from the FIR emission.

Molecular gas observations indicate a gas mass of $9.4\times 10^9\;
\rm M_{\odot}$ while H I observations suggest an atomic gas mass of
$5.5\times 10^9\; M_{\odot}$ (Mirabel \& Sanders 1988).

Our best fit models indicate a total gas mass of about 5.6$\times$
10$^{9}\; \rm M_{\odot}$, of which half is assumed to in the form of
molecular clouds. About 63 percent of the gas has been converted into stars
by the present epoch.

A fit to the SED of IC 5298 is also possible without the inclusion of
the AGN component by assuming a small optical depth for the molecular
clouds, $\tau_{\rm MC}=15$.  This is achieved by decreasing the dust
to gas ratio to 1/300. In this case the best fit to  the FIR and radio
data require that about 25 percent of the mass of the galaxy is formed
during a short intense burst; this galaxy is actually a post--starburst
galaxy.

{\bf UGC 12812} (MRK 0331):   Observational data show evidence of both
starburst (e.g. Lancon, Rocca-Volmerange \& Thuan, 1996; Veilleux et
al., 1995; Roche et al., 1991) and AGN (Lonsdale, Smith \& Lonsdale,
1993) activity.

The 
$23\; \rm GHz$ observation indicates a relatively low thermal
contribution, hence a relatively old starburst, the model also
requires a mild AGN contribution to fit IRAS data.
A younger model, not requiring the AGN contribution,
is marginally compatible with the $23\; \rm GHz$ observation.

Our fits indicate a fair upper limit to the contribution of the AGN
of about 23 percent of the total $8-1000\;\rm \mu m$ luminosity.  The
average SFR is between 72 and $84\; \rm M_{\odot}$/y, for the case
with and without an AGN respectively.

\section{Conclusions}\label{conclusions}

We have performed new high frequency VLA radio measurements ($23\;
\rm GHz$)   of 7  compact ULIRGs chosen from Condon et al.'s
(1991) sample.  We have combined the new observational data with existing
data  at other radio frequencies (1.4 GHz, 5.4 GHz, 8.4 GHz); in 
the millimetre; from SCUBA $850\;\rm \mu m$; infrared IRAS 100, 60 25,
$12\;\rm \mu m$ bands and in the NIR with J, H and K band  observations from
2MASS.

We fit the global SEDs  by means of a combination of new AGN and
starburst models. We find:

\begin{enumerate}

\item The high frequency radio data have proven essential for the 
determination of the relative contribution of the star-burst and the AGN. 
Although AGN
can be detected in several ways (such as via their hard X-ray emission,
near-infrared colours or optical line emission), quantifying the energy
contribution of the AGN to the bolometric luminosity of the source
requires an evaluation of the starburst power. 
On one hand, the new radio observations at 
$23\; \rm GHz$ allow a clean definition (unaffected by free-free absorption) of the shape of
radio emission over a wide range of frequencies. On the other, 
the flux at $23\; \rm GHz$ is mostly contributed ($>$50 percent) by free-free
emission from ionized HII regions and is therefore directly related
to ongoing star formation. The flux at lower frequencies
is instead sensitive to the supernova rate and thus 
carries information about the  
previous history of the starburst.

\item We have found that radio data, together with the $100\;\rm \mu m$ flux
define almost unequivocally the recent star formation
history of the burst. This confirms the suggestion by Bressan, Silva
\& Granato 2002, that combining FIR and Radio data provides a fine
resolution age determination, for these obscured sources.

 \item  While the starburst contribution is generally well constrained by the
SED longward of $60\;\rm \mu m$, the SED below $60\;\rm \mu m$ is sensitive to
the contribution from the hot dust of the AGN. Thus, the excess emission
above the starburst contribution at 12, 25 and $60\;\rm \mu m$ may
be used to quantify the contribution from the AGN. A degree
of uncertainty  remains regarding the strength of the AGN.  Low
optical depth in molecular clouds  favours warm emission from the
starburst, thus a smaller contibution from AGN is required to fit the
mid-infrared data. However, as the emission from the starburst gets
warmer, its contribution at $100\;\rm \mu m$ decreases and to fit this
data point a larger SFR is required; eventually becoming incompatible
with the radio data. Thus, while still affected by uncertainties, the
method we propose proves to be useful in providing an estimate of the
relative contribution of AGN and starburst.

\item The effect of the AGN is also visible from the total J, H and K fluxes 
but its quantification is impossible from these NIR colours alone
because of the strong dependence of the AGN emission on the model
parameters at these wavelengths.

\item It is possible to fit the SEDs of most of the sources in our sample
with no contribution from an AGN. However, better fits, especially for the
near infrared colours, are obtained when the bolometric luminosity has a
small AGN contribution. We note that Farrah et al. (2003), who fitted the
SEDs of ULIRGs with template spectra, find that in general the starburst
contributions dominante.

\item To fit the mid-infrared data a very high optical depth at 1 micron is
generally required. In the most extreme case, IRAS01364-1042, the SED cannot
be fit unless $\tau_{1\;\rm \mu m}\sim 180$. Although we cannot do an object
by object comparison, values as high as at least 50 are consistent with gas
column densities measured in the centre of ULIRGs. The fact that extinction
measurements at shorter wavelengths give much lower values probably indicates
that such observations penetrate only a surface layer of the ISM and do not
probe the whole sight-line towards the central energy source.

\end{enumerate}

\begin{acknowledgements}
We thank Greg Taylor, Claire Chandler and Ylva Pihlstroem for
invaluable advice regarding the observations and data reduction. 
A.B. and G.L.G. acknowledge warm hospitality by INAOE and discussions
with I. Aretxaga, O. Vega, D. Hughes, T.N.  Rengarajan    and
R. Terlevich.  We thank the referee, M. Rowan-Robinson, for useful
suggestions.
This research was partially supported by the European
Commission Research Training Network `POE' under contract
HPRN-CT-2000-00138 and by the Italian Ministry for University and
Research (MURST) under grant Cofin 92001021149-002. The National Radio
Astronomy Observatory is a facility of the National Science Foundation
operated under cooperative agreement by Associated Universities,
Inc. This publication makes use of data products from the Two Micron
All Sky Survey, which is a joint project of the University of
Massachusetts and the Infrared Processing and Analysis
Center/California Institute of Technology, funded by the National
Aeronautics and Space Administration and the National Science
Foundation.

\end{acknowledgements}

\end{document}